\documentclass[10pt]{article}
\usepackage{graphicx}
\usepackage{lineno}

\def\Title#1{\begin{center} {\Large #1 } \end{center}}
\def\Author#1{\begin{center}{ \sc #1} \end{center}}
\def\Address#1{\begin{center}{ \it #1} \end{center}}

\newcommand\pubblock{\rightline{\begin{tabular}{l} Proceedings of the Fifth Annual LHCP\\ \pubnumber\\
         \pubdate  \end{tabular}}}

\newenvironment{Abstract}{\begin{quotation} \begin{center} 
             \large ABSTRACT \end{center}\bigskip 
      \begin{center}\begin{large}}{\end{large}\end{center} \end{quotation}}

\newenvironment{Presented}{\begin{quotation} \begin{center} 
             PRESENTED AT\end{center}\bigskip 
      \begin{center}\begin{large}}{\end{large}\end{center} \end{quotation}}

\def\beq{\begin{equation}}
\def\eeq#1{\label{#1}\end{equation}}
\def\eeqn{\end{equation}}

\def\beqa{\begin{eqnarray}}
\def\eeqa#1{\label{#1}\end{eqnarray}}
\def\eeqan{\end{eqnarray}}

\let\bar=\overbar

\def\Dslash{\not{\hbox{\kern-4pt $D$}}}
\def\dslash{\not{\hbox{\kern-2pt $\del$}}}

\def\msb{{\bar{\ssstyle M \kern -1pt S}}}

\usepackage{color}

\newcommand\pubnumber{ ATL-PHYS-PROC-2017-064 }

\newcommand\pubdate{June 9, 2017}

\def\affiliation{
On behalf of the ATLAS Experiment, \\
Institute of Nuclear Physics, Polish Academy of Sciences \\
Krakow, Poland}

\def\support{\footnote{Work supported in part by the National Science Centre, Poland, grant 2015/18/M/ST2/00087 and
		by PL-Grid Infrastructure. }}

\begin{document}

\large
\begin{titlepage}
\pubblock

\vfill
\Title{  NEW RESULTS ON COLLECTIVITY WITH ATLAS  }
\vfill

\Author{ KRZYSZTOF W. WO\'{Z}NIAK \support }
\Address{\affiliation}
\vfill

\begin{Abstract}

The collective phenomena are observed not only in heavy ion collisions, but also in the proton-nucleus and in high-multiplicity 
$pp$ collisions. The latest results from this area obtained in 
ATLAS are presented. In $p$+Pb collisions the emission source of particles is measured using the HBT method. The analysis of $p$+Pb data collected in 2016 provides information on the elliptic flow of charged hadrons and muons. Low multiplicity events from $pp$, $p$+Pb and peripheral Pb+Pb collisions are studied with the cumulant methods. A deeper understanding of Pb+Pb collisions is provided by the analysis of longitudinal fluctuations of the collective flow parameters.

\end{Abstract}
\vfill

\begin{Presented}
The Fifth Annual Conference\\
 on Large Hadron Collider Physics \\
Shanghai Jiao Tong University, Shanghai, China\\ 
May 15-20, 2017
\end{Presented}
\vfill
\end{titlepage}
\def\thefootnote{\fnsymbol{footnote}}
\setcounter{footnote}{0}
%

\normalsize 


\begin{figure}[t]
\centering
\includegraphics[width=0.51\textwidth]{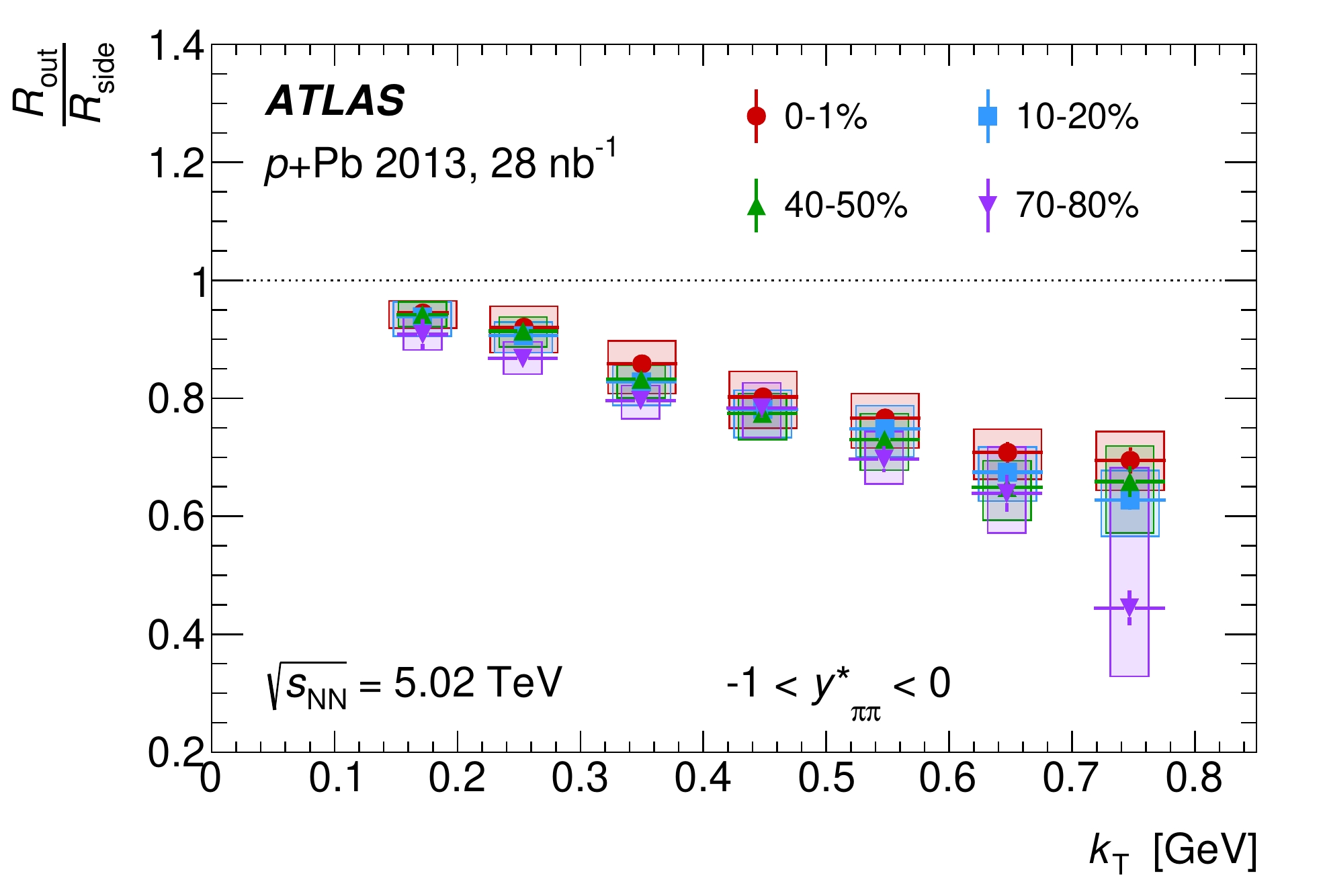}
\includegraphics[width=0.48\textwidth]{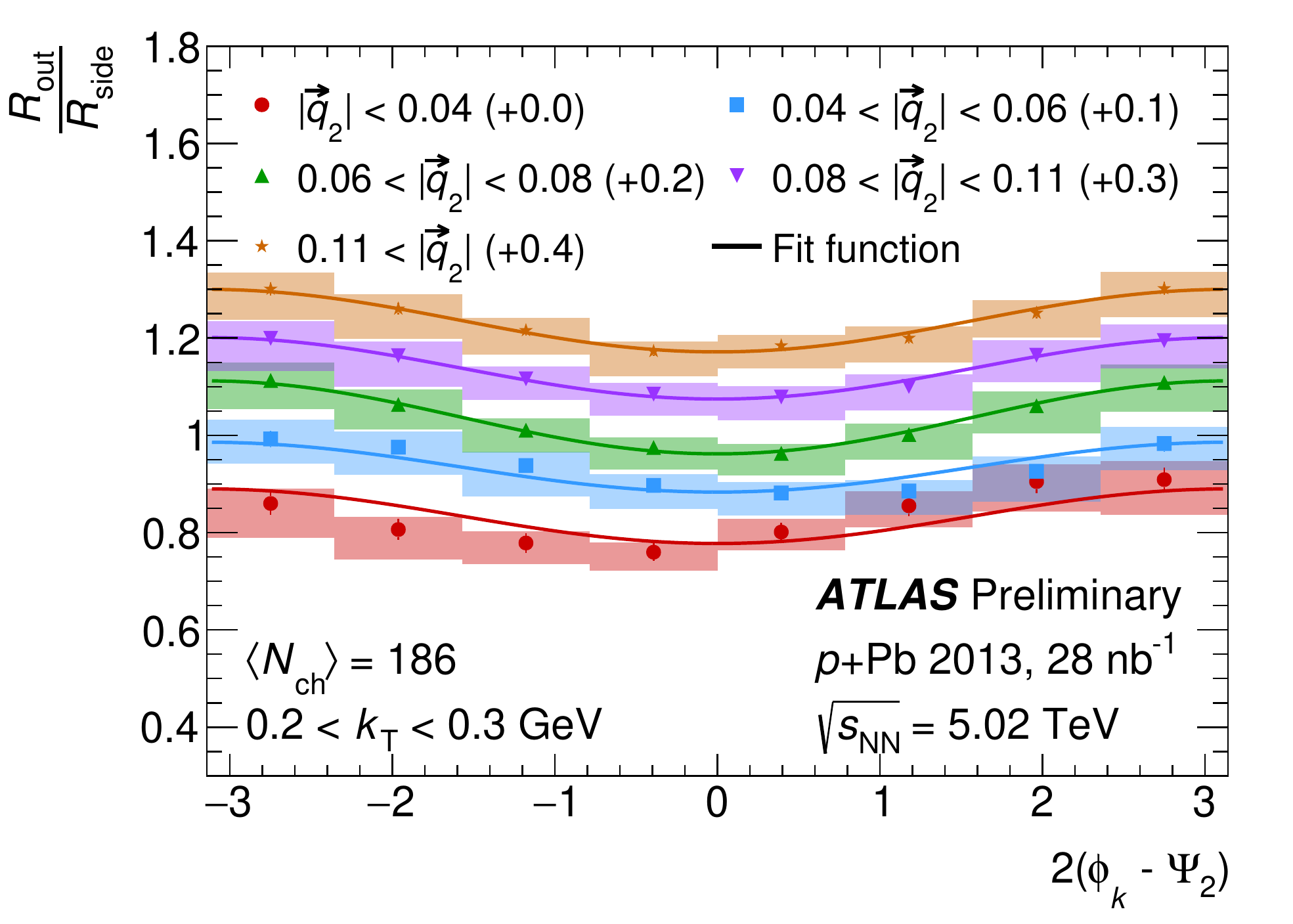}
\vspace*{-0.5cm}
	\caption{ The ratio of exponential radii  $R_{\mathrm{out}}/R_{\mathrm{side}}$ (left) 
as a function of the pair transverse momentum, 
$k_{\mathrm{T}}$, 
in four centrality bins \cite{atlas:femto_01621}
	and (right) as a function of the angular distance from 
the second-order event plane $2(\phi_{k}-\Psi_{2})$ in five  
bins of  elliptic flow vector magnitude $|\vec{q}_{2}|$~\cite{atlas:femto_008}.
The points for different 
intervals of $|\vec{q}_{2}|$ are offset for 
visibility, as indicated in the legend. 
	}
\label{fig:figure1}
\end{figure}

\section{Introduction}

Strong collective effects were first found in heavy ion collisions, where they indicate a 
creation of the Quark-Gluon Plasma. This phase of matter manifests azimuthal correlations among produced particles, referred to as collective particle flow. However, similar effects are observed in proton-nucleus and even in high-multiplicity proton-proton collisions.
The measurements of Pb+Pb, $p$+Pb and $pp$ collisions at energies available from the Large Hadron Collider (LHC) and performed using 
the ATLAS detector~\cite{atlas:detector} allow to study collective phenomena in detail. The obtained results are important for understanding the particle production at LHC energies.

\section{Results}

One of the questions, which can be answered by analysing correlations
among produced particles is the size of their emission source. 
The Bose-Einstein correlations of identical bosons, 
as a function of the difference of their momenta, 
${\mathrm{\bf q}} = {\mathrm{\bf p^{a}}} - {\mathrm{\bf p^{b}}}$, 
can be parameterized as:
\begin{equation}
C_{\mathrm{BE}}({\bf q}) = 1 + e^{\|{\bf Rq}\|},
\end{equation}
where the diagonal elements of the matrix ${\bf R}$ are $R_{\mathrm{out}}$, $R_{\mathrm{side}}$ and $R_{\mathrm{long}}$. 
These parameters were studied for $p$+Pb collisions at $\sqrt{s_{_{\mathrm{NN}}}} = 5.02$~TeV~\cite{atlas:femto_01621} 
as a function of $k_{\mathrm{T}}$ 
(where $\mathrm{\bf k} = 0.5 (\mathrm{\bf p^{a}} - \mathrm{\bf p^{b}})$) and the number of nucleons participating in collisions, $N_{\mathrm{part}}$.
Especially interesting is the ratio $R_{\mathrm{out}} / R_{\mathrm{side}}$ which represents asymmetry of the source. 
In Figure~\ref{fig:figure1}~(left) one can see that this ratio 
depends on $k_{\mathrm{T}}$, which indicates radial expansion of 
the source of particles~\cite{atlas:femto_01621}. This ratio depends also on the orientation of 
the correlated pairs (represented by $\phi_{k}$, azimuthal angle of  $k_{\mathrm{T}}$) with respect to the event plane, $\Psi_{2}$,
as shown in Figure~\ref{fig:figure1}~(right).
The  $R_{\mathrm{out}} / R_{\mathrm{side}}$ ratio is larger out of plane than in plane 
($\phi_{k}\approx \Psi_{2}$) as $R_{\mathrm{out}}$ is enhanced out of plane and $R_{\mathrm{side}}$ does not change much with 
$\phi_{k}-\Psi_{2}$~\cite{atlas:femto_008}.

\begin{figure}[t]
	\centering
	\includegraphics[width=0.49\textwidth]{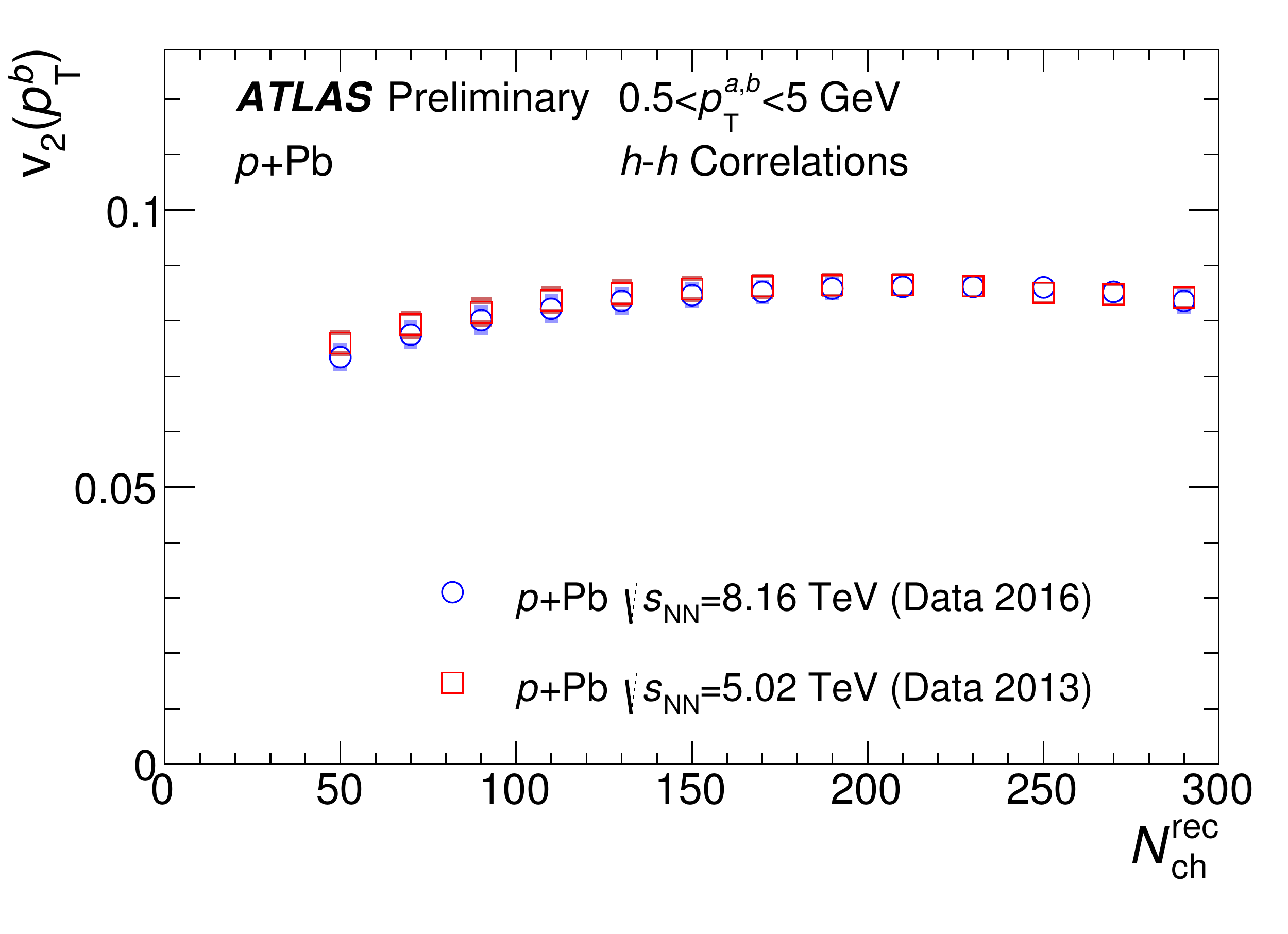}
	\includegraphics[width=0.49\textwidth]{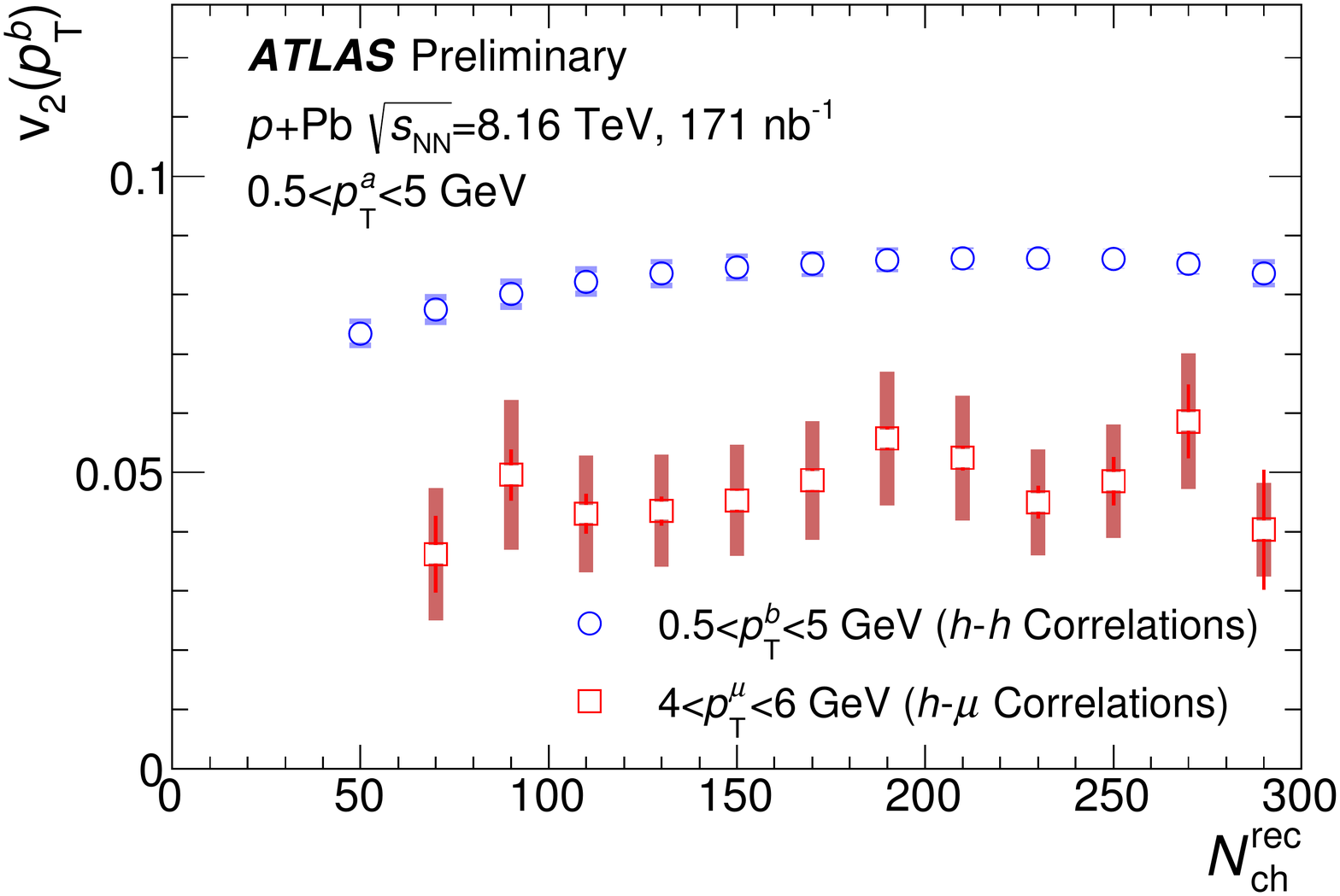}
	\vspace*{-0.2cm}
	\caption{ The $v_{2}$ values, as a function of $N_{\mathrm{ch}}^{\mathrm{rec}}$,  obtained (left) in $p$+Pb collisions at 8.16 TeV and 5.02 TeV
		and (right) in $p$+Pb collisions at 8.16 TeV 
from $h$-$h$ 
correlations (circles) 
and $h$-$\mu$ correlations
		(squares)~\cite{atlas:muons_006}.
	}
	\label{fig:figure2}
\vspace*{0.4cm}
	\centering
	\includegraphics[width=0.46\textwidth]{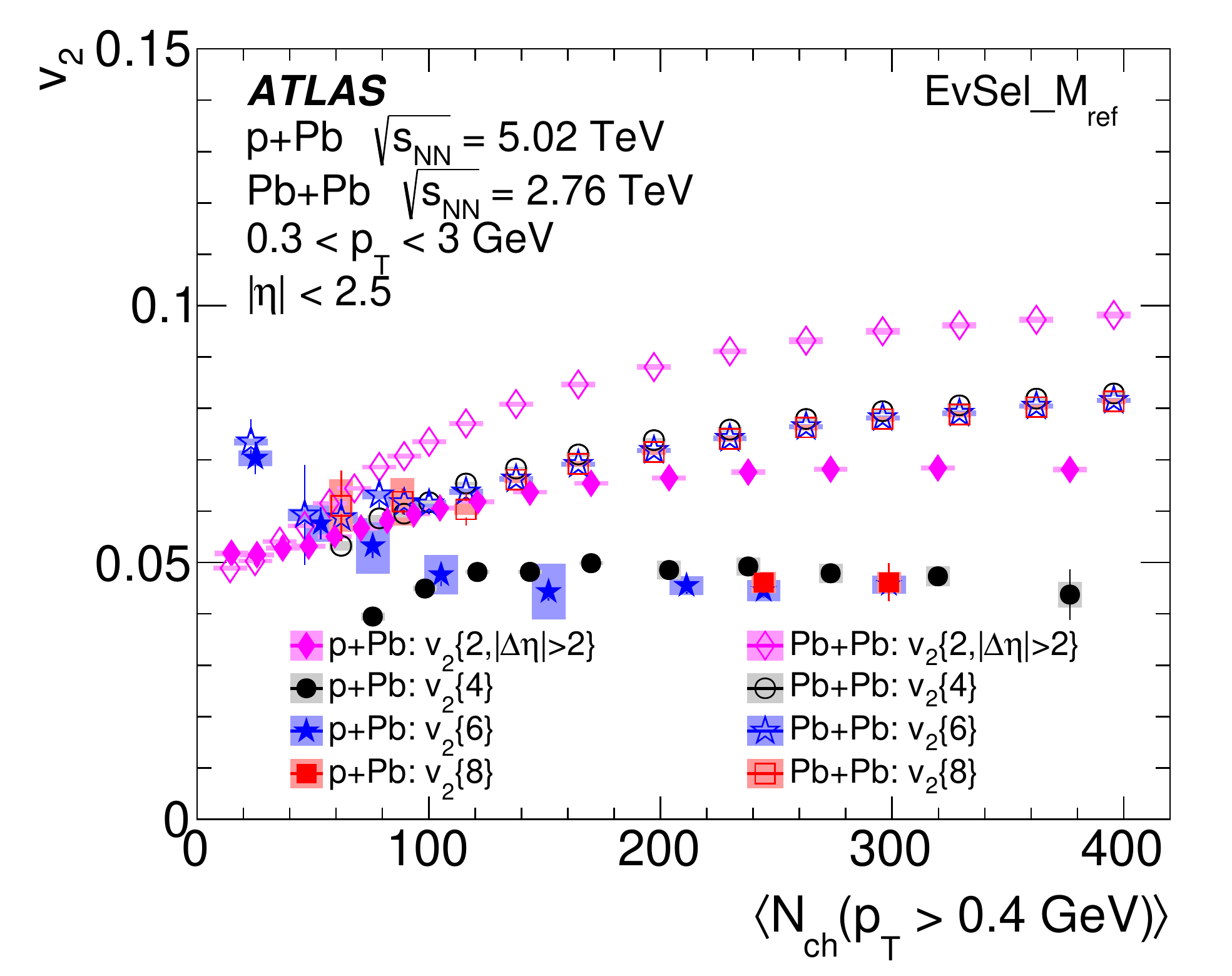}
	\includegraphics[width=0.53\textwidth]{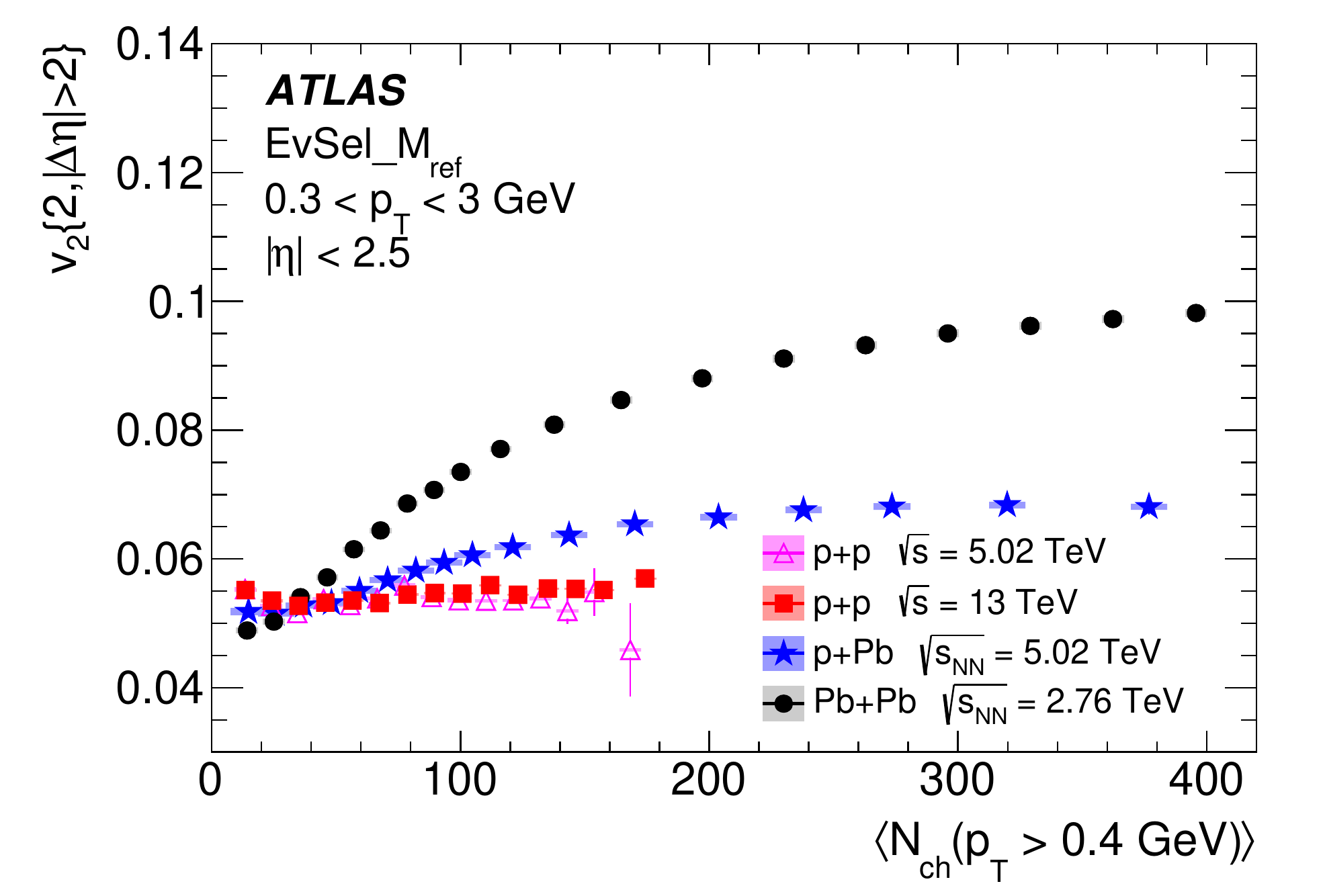}
	\vspace*{-0.4cm}

	\caption{ Multiplicity dependence  of (left)
	 $v_{2}\{2,|\Delta\eta| > 2\}$,
	 $v_{2}\{4\}$, $v_{2}\{6\}$ and $v_{2}\{8\}$ for $p$+Pb collisions at 
	 5.02~TeV and low-multiplicity Pb+Pb collisions at 2.76~TeV 
	and (right) $v_{2}\{2,|\Delta\eta| > 2\}$ for $pp$ collisions
	 at 5.02~TeV and 13~TeV, 
	 $p$+Pb collisions at  5.02~TeV
	 and low-multiplicity Pb+Pb collisions at  2.76~TeV~\cite{atlas:cumul_007}.
	}
	\label{fig:figure3}
	\vspace*{0.3cm}
\end{figure}

\begin{figure}[p]
	\centering
	\includegraphics[width=0.98\textwidth]{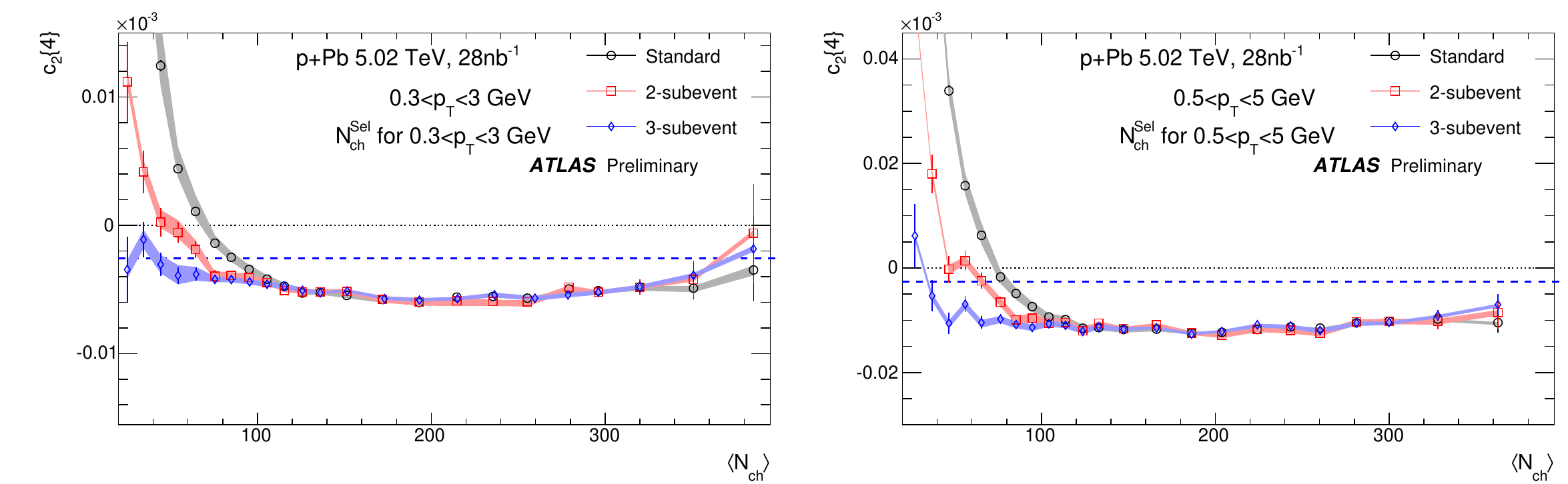}
	\vspace*{-0.2cm}
	\caption{ The $c_{2}\{4\}$ cumulants in $p$+Pb collisions 
		at 5.02~TeV calculated for charged particles 
with (left) 
		$0.3<p_{\mathrm{T}}<3$~GeV and (right)
		$0.5<p_{\mathrm{T}}<5$~GeV compared across 
the three cumulant 
methods~\cite{atlas:cumul_subev_002}. 
	}
	\label{fig:figure5}
	\vspace*{0.4cm}

	\centering
	\includegraphics[width=0.98\textwidth]{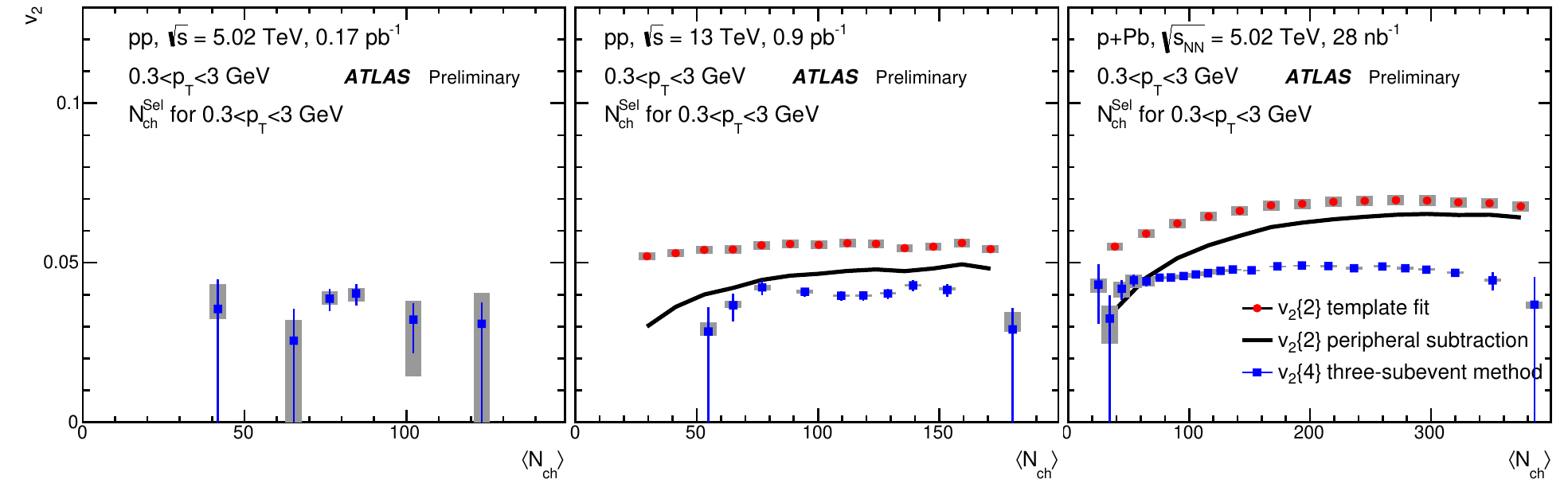}
	\vspace*{-0.2cm}
	\caption{ The $v_{2}\{4\}$ values calculated for charged 
		particles with 	$0.3<p_{\mathrm{T}}<3$~GeV using 
		the three-subevent method in 
		$pp$ collisions at (left) 5.02~TeV, (middle) 13~TeV
		and (right) in $p$+Pb collisions at 5.02~TeV.
		They are compared to $v_{2}$ obtained from a two-particle correlation analysis using a template fit procedure 
		(solid circles) or peripheral subtraction (solid line)
		to remove non-flow effects~\cite{atlas:cumul_subev_002}. 
	}
	\label{fig:figure6}
\vspace*{0.4cm}

	\centering
	\includegraphics[width=0.98\textwidth]{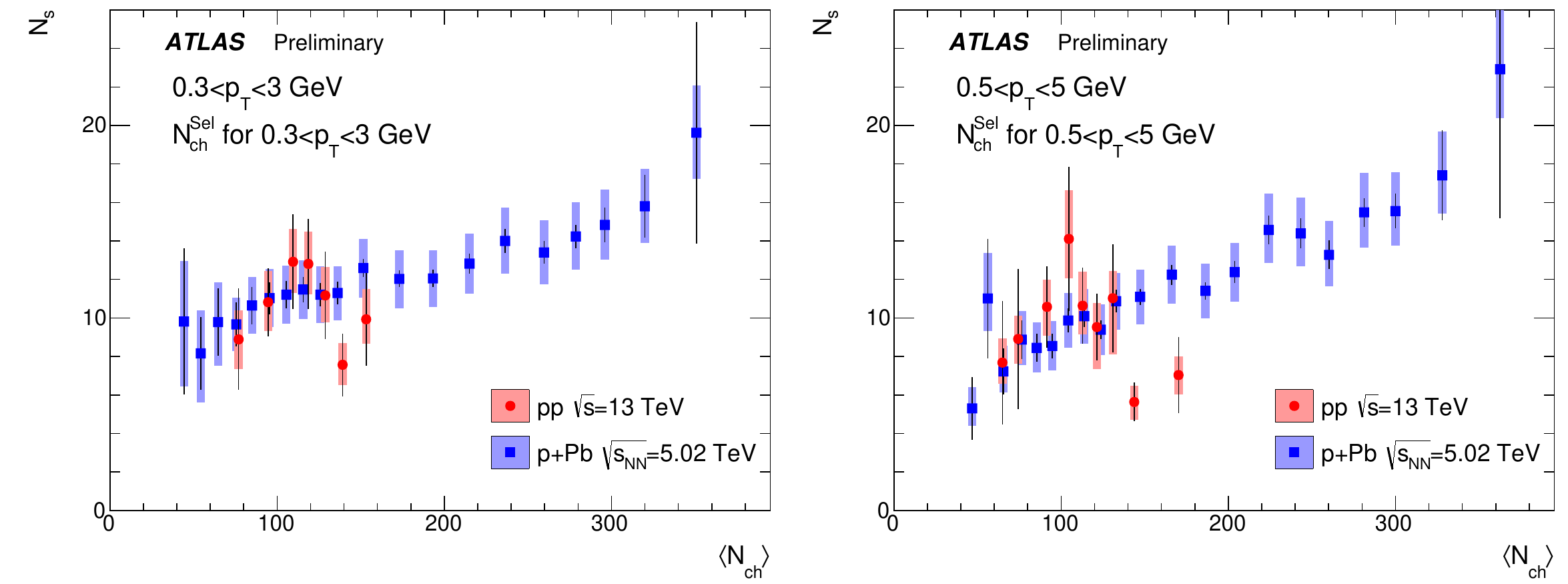}
	\vspace*{-0.2cm}
	\caption{ The number of sources inferred from $v_{2}\{2\}$ and $v_{2}\{4\}$ using Eq.~\ref{equ:cumul_Ns} in $pp$ and 
		$p$+Pb collisions at 13~TeV and 5.02~TeV, respectively, 
		as a function of charged-particle multiplicity, 
		$N_{\mathrm{ch}}^{\mathrm{Sel}}$, obtained by selecting 
		charged particles with (left) $0.3<p_{\mathrm{T}}<3$~GeV   
	        and (right) 
		$0.5<p_{\mathrm{T}}<5$~GeV~\cite{atlas:cumul_subev_002}. 
	}
	\label{fig:figure7}
\end{figure}

In the studies of azimuthal correlations among charged particles
 the Fourier decomposition is used:
\begin{equation}
\frac{ \mathrm{d}N_{\mathrm{ch}} }{ \mathrm{d}\phi } \sim 1 + 2 \sum_{n=1}^{\infty} 
v_{n}(p_{\mathrm{T}}, \eta) \cos(n(\phi-\Psi_{n})).
\end{equation}
The flow harmonics $v_{n}$ can be obtained also from
two-particle correlations. In $pp$ and $p$+Pb collisions these correlations contain large
contributions  from other sources (non-flow), which
have to be subtracted. ATLAS applies a {\it template method}~\cite{atlas:template_prl116} to remove them. 
In Figure~\ref{fig:figure2} the elliptic flow $v_{2}$ in $p$+Pb collisions at 5.02~TeV and 
8.16~TeV for all hadrons ($h$-$h$) and for 
hadron-muon ($h$-$\mu$) correlations (at~8.16~TeV) is shown as 
a function of 
charged-particle multiplicity, 
$N_{\mathrm{ch}}^{\mathrm{rec}}$~\cite{atlas:muons_006}. 
There is no dependence of $v_{2}$ on the energy
of collisions and a very weak increase with multiplicity. 
The values of $v_{2}^{h-\mu}$ reach only about 
$0.6\, v_{2}^{h-h}$.

The non-flow effects are most pronounced in low-multiplicity 
events as they usually involve relatively small number 
of particles in limited kinematical range (resonance decays, jets)
and are thus suppressed in multi-particle correlations. 
Such correlation can be used to calculate cumulants, 
$c_{n}\{2k\}$~\cite{atlas:cumul_007}, closely related to
flow harmonics: 
\begin{equation}
v_{n}\{2\} = \sqrt{c_{n}\{2\}}, ~~~~~~~ 
v_{n}\{4\} = \sqrt[4]{-c_{n}\{4\}}, ~~~~~ 
v_{n}\{6\} = \sqrt[6]{-c_{n}\{6\}/4}, ~~~~~
v_{n}\{8\} = \sqrt[8]{-c_{n}\{8\}/33}.
\label{equ:cumul_vn}
\end{equation}
In the case of $c_{n}\{2\}$ a separation of particles in pseudorapidity ($|\Delta\eta|>2$) is required 
in calculations of $v_{n}\{2, |\Delta\eta|>2\}$ 
to suppress short-range correlations.  
In Figure~\ref{fig:figure3} elliptic flow $v_{2}$ 
values obtained for $p$+Pb, Pb+Pb and $pp$ collisions 
using different multi-particle cumulants are compared~\cite{atlas:cumul_007}. 
While the $v_{2}\{2k\}$
values are similar (for k=2, 3, 4), those for $k=1$ 
with $|\Delta\eta|>2$ requirement are larger. Comparison
of \mbox{$v_{2}\{2, |\Delta\eta|>2\}$} for different systems 
shows
an increase with the size of colliding projectiles  
for events with the same number of produced particles.
The $v_{2}\{2k\}$ harmonics do not change with the multiplicity 
and the energy for $pp$ collisions while are increasing 
with multiplicity for $p$+Pb and Pb+Pb collisions.

Further suppression of short range correlations can be 
achieved if in the calculations of correlations, used to 
obtain cumulants,  particles from different ranges of
pseudorapidity (i.e. {\it subevents})
are used~\cite{atlas:cumul_subev_002}. For $c_{2}\{4\}$
negative values are expected, as otherwise $v_{2}\{4\}$ 
can not be calculated using Eq.~\ref{equ:cumul_vn}.
In Figure~\ref{fig:figure5} one can see that positive 
$c_{2}\{4\}$
are obtained at low multiplicities, but are reduced
in two- and especially three-subevent cumulant methods. 
The  $v_{2}\{4\}$ values from three-subevent method are 
lower than $v_{2}\{2\}$ from peripheral subtraction or 
template fit method (Figure~\ref{fig:figure6}). 
This difference can be interpreted
as a result of event-by-event flow fluctuations,  
which are closely related to the effective number of sources, $N_{s}$, for particle production:
\begin{equation}
\frac{v_{2}\{4\}}{v_{2}\{2\}}  = 
\left(\frac{4}{3+N_{s}}\right)^{1/4}.
\label{equ:cumul_Ns}
\end{equation}
The number of sources shown in Figure~\ref{fig:figure7} 
is similar for $pp$ and $p$+Pb collisions and for the latter it increases from 10 to 20
in the full available event-multiplicity range.

\begin{figure}[t]
	\centering
	\includegraphics[width=0.49\textwidth]{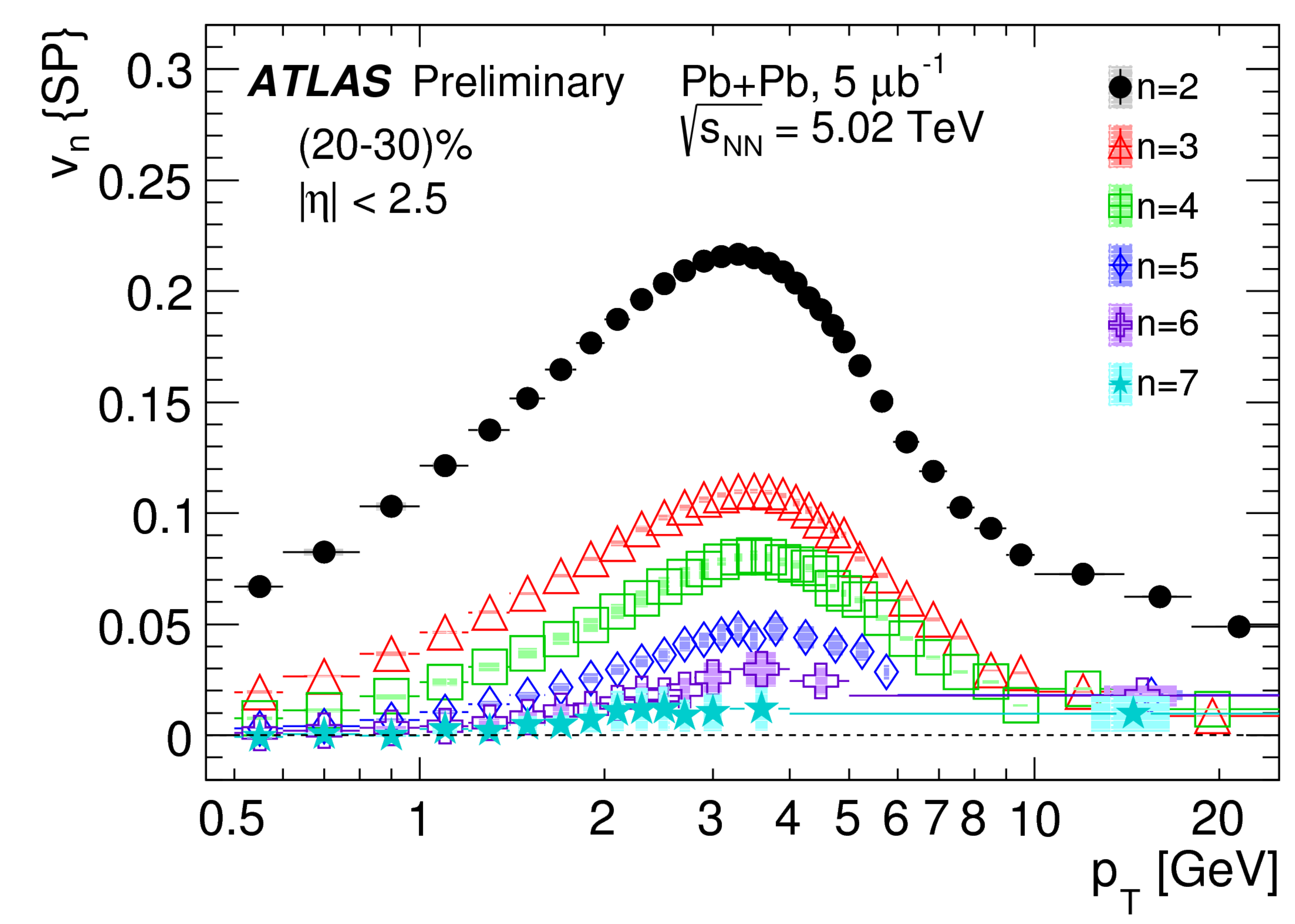}
	\includegraphics[width=0.49\textwidth]{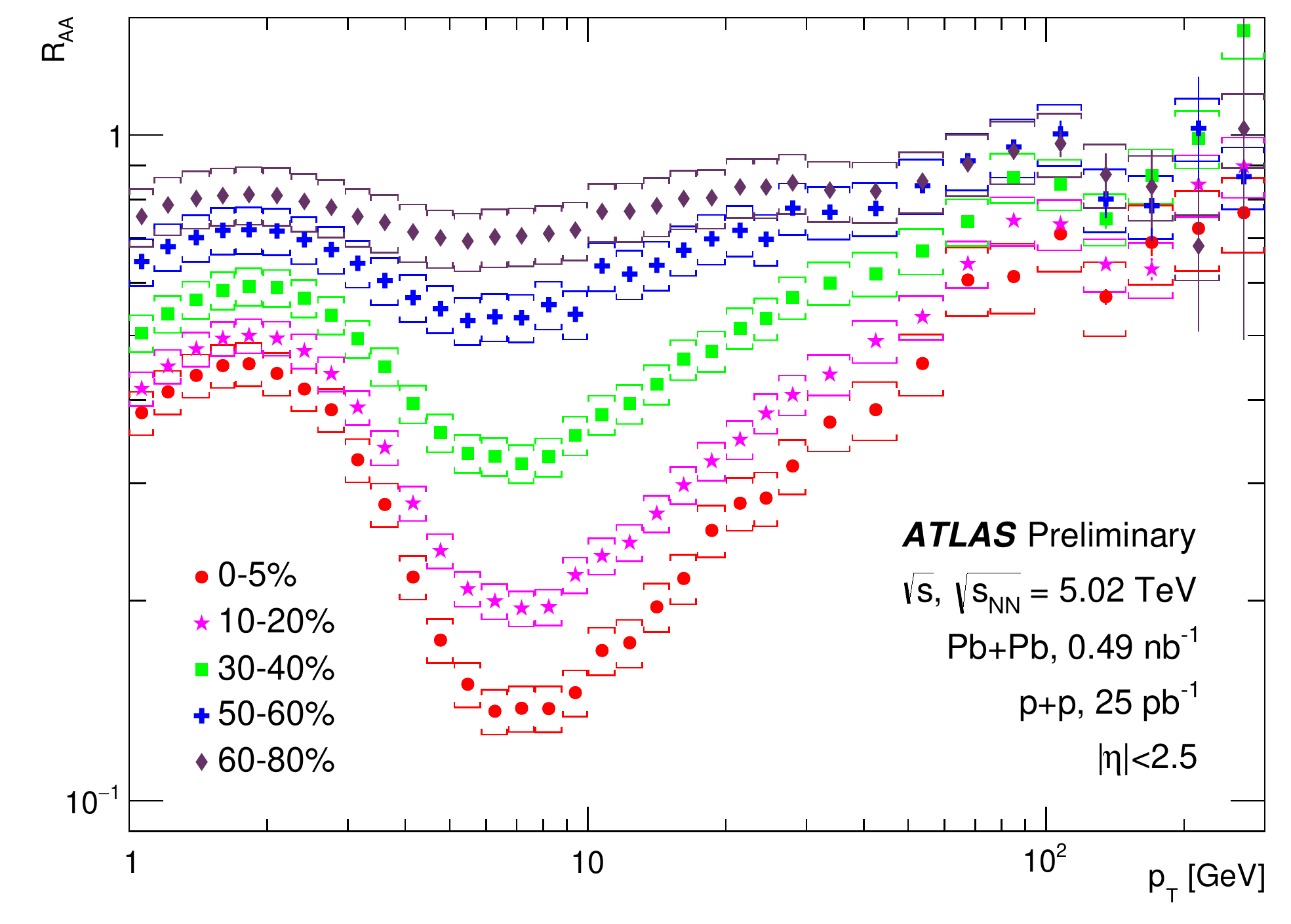}
	\vspace*{-0.2cm}
	\caption{ (left) The $v_{n}$ values as a function of transverse momentum $p_{\mathrm{T}}$, 
        in (20--30)\% centrality interval 
	of Pb+Pb collisions at 5.02~TeV~\cite{atlas:vn_pbpb_105}.
	(right) Nuclear modification factor $R_{\mathrm{AA}}$ as a function of $p_{\mathrm{T}}$ 
        for charged hadrons measured 
	in Pb+Pb collisions at 5.02~TeV for five centrality
	intervals~\cite{atlas:raa_pbpb_012}.
	}
	\label{fig:figure8}
\vspace*{0.4cm}

	\centering
	\includegraphics[width=0.49\textwidth]{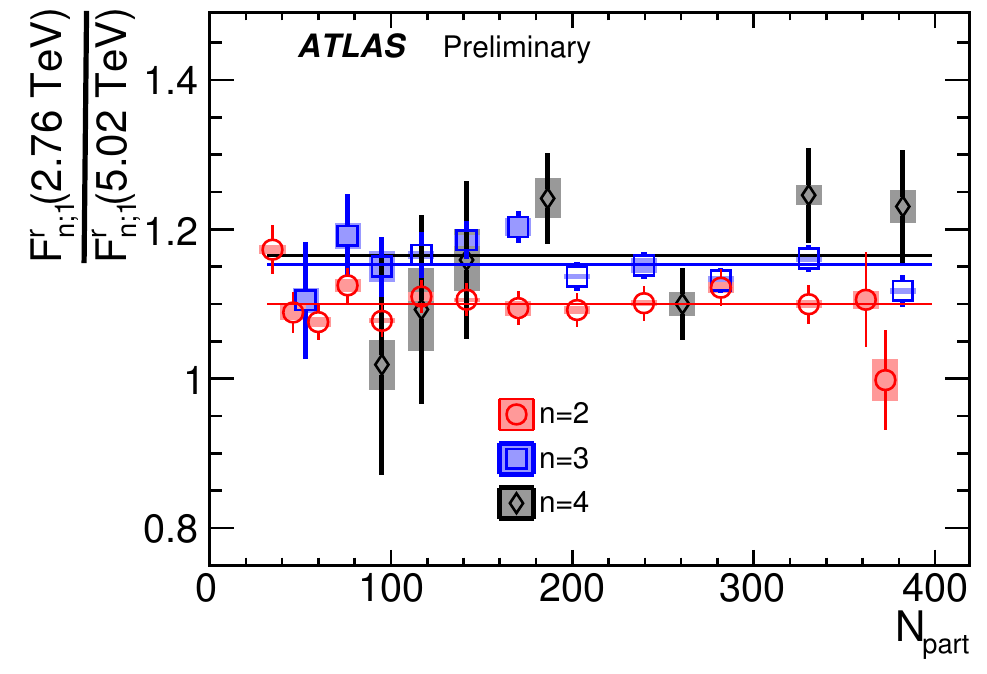}
	\includegraphics[width=0.49\textwidth]{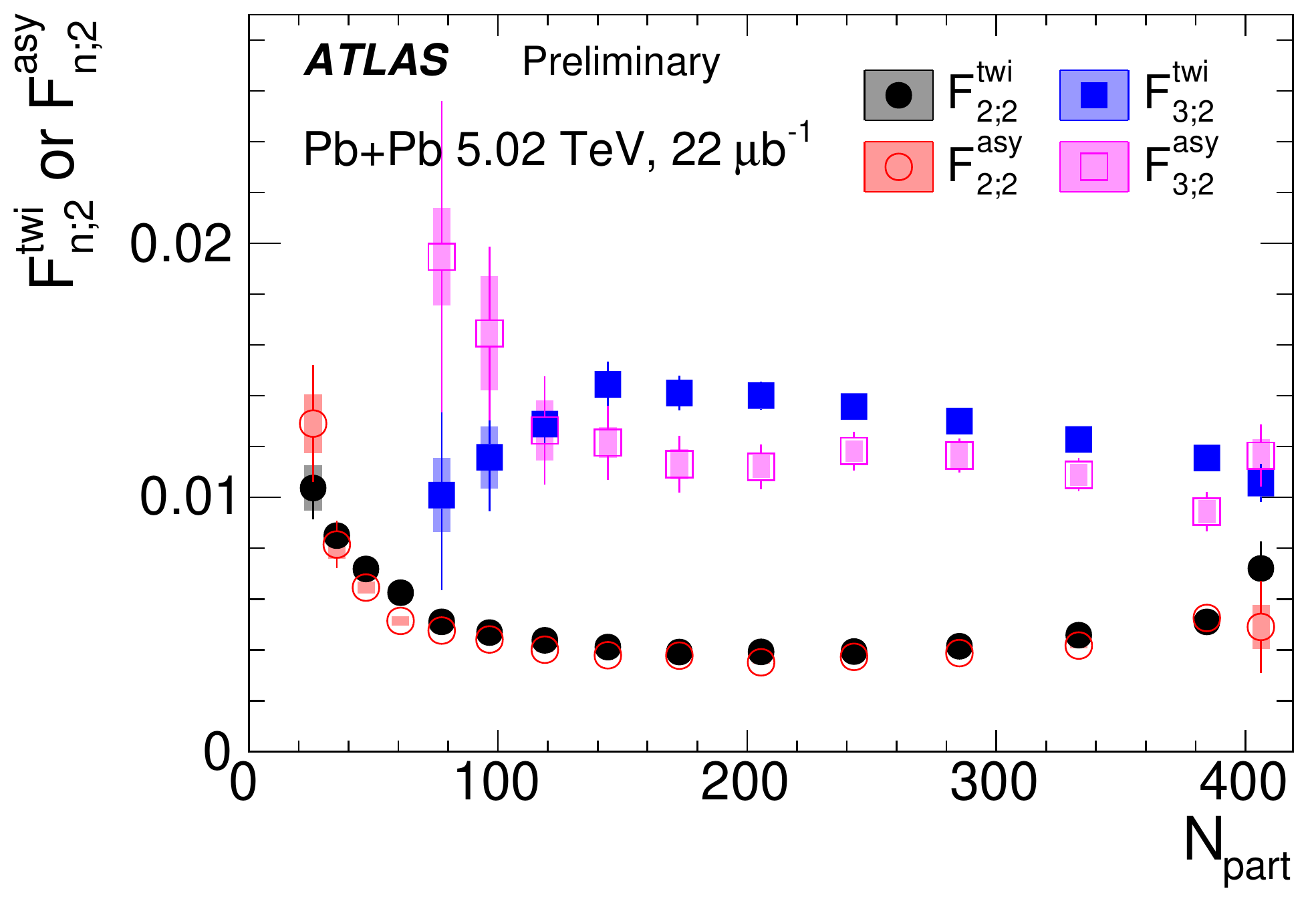}
	\vspace*{-0.2cm}
	\caption{ (left) The ratio of decorrelation parameter
	$F_{\mathrm{n;1}}^{\mathrm{r}}$ in Pb+Pb collisions at 
	2.76 TeV and 5.02 TeV as a function of centrality represented 
	by $N_{\mathrm{part}}$.
	(right) The event-plane twist component $F_{n;2}^{\mathrm{twi}}$ and asymmetry 
        component $F_{n;2}^{\mathrm{asy}}$ as a function of $N_{\mathrm{part}}$ 
        for $n$=2 and $n$=3 for Pb+Pb collisions at 5.02~TeV~\cite{atlas:decorrel_003}. 
	}
	\label{fig:figure9}
\end{figure}

For events with large multiplicities, such as measured 
in Pb+Pb collisions at 5.02~TeV, standard methods of 
calculation of flow harmonics are sufficiently robust against non-flow effects. 
In Figure~\ref{fig:figure8}~(left) $v_{n}$, for $n=2-7$,
as a function of $p_{\mathrm{T}}$ in 20-30\% centrality
interval are shown~\cite{atlas:vn_pbpb_105}.
Flow harmonics up to $v_{7}$ are \mbox{non-zero}. The $v_{2}$ 
is 0.05 
even for $p_{\mathrm{T}} > 20$~GeV. Relatively large $v_{2}$ for
particles with high transverse momenta means that also very energetic 
partons are interacting in the QGP. This observation is consistent
with the measured suppression of high-$p_{\mathrm{T}}$ 
charged hadrons as quantified by the nuclear modification 
factor, 
$R_{\mathrm{AA}}$~\cite{atlas:raa_pbpb_012}. 
In Figure~\ref{fig:figure8}~(right) one can see that 
$R_{\mathrm{AA}}<1$ in the same centrality 
and $p_{_{\mathrm{T}}}$ range.

A deeper understanding of flow phenomena may provide study of  the dependence of flow fluctuations on position in pseudorapidity using correlators $r_{n|n;k}$ and $R_{n,n|n,n}$ 
(see Ref.~\cite{atlas:decorrel_003} for definitions).
Assuming that their dependence on pseudorapidity is linear it 
can be parameterized as:
\begin{equation} 
r_{n|n;k} \approx 1 - 4\, k\, F_{n;k}^{\mathrm{r}}\, \eta
~~~~~~ \mathrm{and} ~~~~~~
R_{n,n|n,n} \approx  1 - 4\, F_{n;n}^{\mathrm{twi}}\, \eta,
\end{equation}
where 
$F_{n;k}^{\mathrm{r}}=F_{n;n}^{\mathrm{asy}}+F_{n;n}^{\mathrm{twi}}$,
 $F_{n;n}^{\mathrm{asy}}$ and $F_{n;n}^{\mathrm{twi}}$
are decorrelation parameters connected with magnitude (asymmetry) 
and twist fluctuations, respectively. In Pb+Pb collisions the decorrelation
parameters $F_{n;k}^{\mathrm{r}}$ are similar for all centralities
but depend on the energy of the collision 
and are 10-16\% larger at 2.76~TeV than at 5.02~TeV 
(see Figure~\ref{fig:figure9}~(left)). 
On the other hand the magnitude and twist decorrelation parameters,
shown in Figure~\ref{fig:figure9}~(right),
are approximately constant for $N_{\mathrm{part}}>100$.
In the whole multiplicity range  $F_{n;2}^{\mathrm{asy}} \approx F_{n;2}^{\mathrm{twi}}$ for the same $n$~\cite{atlas:decorrel_003}.

\section{Conclusions}

Studies of correlations among particles produced in different types
of collisions available at LHC provide valuable information 
on properties of their source.
In $p$+Pb collisions the volume from which particles are emitted has an elongated shape and undergoes a radial expansion. The flow of
muons originating from $b$ or $c$ quarks is much smaller than 
that of charged hadrons. Results on flow harmonics obtained using
cumulant methods clearly show that   
the non-flow contributions are very important in low-multiplicity events and need to be properly subtracted. 
In new detailed studies of Pb+Pb collisions non-zero flow 
harmonics
up to $v_{7}$ were measured. Analysis of longitudinal 
fluctuations of 
flow harmonics reveals decorrelation effects which are stronger at lower collision energy, but similar when decomposed into magnitude
and twist contributions.


\end{document}